\documentclass[12pt,a4paper]{article}
\usepackage{dfttrob}
\usepackage[numbers,sort&compress]{natbib}
\usepackage{latexuseful2e}
\usepackage{amsmath}
\usepackage{amssymb}
\usepackage{graphicx}

\dfttnum{DFTT 21/2003\\December 2003}

\newcommand{\III}{\ensuremath{\text{th}}}
\newcommand{\nbarsoft}{\nbar_{\text{soft}}}
\newcommand{\nbarsemi}{\nbar_{\text{sh}}}
\newcommand{\nbariii}{\nbar_{\III}}
\newcommand{\ksoft}{k_{\text{soft}}}
\newcommand{\ktotal}{k_{\text{total}}}
\newcommand{\ksemi}{k_{\text{sh}}}
\newcommand{\kiii}{k_{\III}}
\newcommand{\Nbarsoft}{\Nbar_{\text{soft}}}
\newcommand{\Nbarsemi}{\Nbar_{\text{sh}}}

\newcommand{\ncsemi}{\nbar_{c,\text{sh}}}

\newcommand{\alphasoft}{\alpha_{\text{soft}}}

\newcommand{\rs}{\sqrt{s}}
\newcommand{\ktot}{k_{\text{total}}}

\newcommand{\bFB}{b_{\text{FB}}}
\newcommand{\FPS}{{\text{FPS}}}

\begin{document}

\title{Signals of new physics in global event properties in pp
	collisions in the TeV energy domain: rapidity intervals}
\author{A. Giovannini and R. Ugoccioni\\
 \it Dipartimento di Fisica Teorica and INFN - Sez. di Torino\\
 \it Via P. Giuria 1, 10125 Torino, Italy}
\maketitle

\begin{abstract}
The study of possible new physics signals in global event properties
in pp collisions in the TeV energy domain is extended from full
phase-space to rapidity intervals experimentally accessible at LHC.
The elbow structure in the total multiplicity distribution predicted
in full phase-space is clearly present also in restricted rapidity
intervals,  leading to very strong charged particle correlations.
It is also found that energy densities comparable to those reached in
heavy ion collisions at RHIC could be attained in pp collisions at LHC.
\end{abstract}

\section*{Introduction}
The search for signals of new physics in global event properties in pp
collisions at LHC in the framework of the weighted superposition
mechanism of different classes of minimum bias events as components led
us to explore the possibility of the existence at 14 TeV c.m.\ energy
of a new class of events \cite{RU:NewPhysics}.
The aim of the present paper is to extend our study from full
phase-space (FPS) to pseudo-rapidity intervals which will become
experimentally accessible with the LHC detectors.

The paper is organised as follows.
In Section 1 are summarised results of \cite{combo:eta} on general properties of
soft and semi-hard components in pseudo-rapidity intervals in
extrapolated scenarios based on the knowledge of the GeV energy
domain.
In Section 2 the existence at 14 TeV c.m.\ energy in pp collisions of
a third class of events in addition to the soft and semi-hard ones is
postulated following our findings  on the same class in FPS and its
general properties in pseudo-rapidity intervals are discussed.
For completeness, a comparison with Pythia Monte Carlo calculation
results is performed where appropriate.

\section{Antecedents: global properties in full phase-space}
In \cite{combo:eta} the total $n$-charged particle multiplicity
distribution in pseudo-rapidity intervals $|\eta|<\eta_c$ as a
function of c.m.\ energy $\sqrt{s}$ was written as follows:
\begin{multline}
	P_n(\eta_c,\rs) = 
	   \alphasoft(\rs) P_n^{\text{PaNB}}(\nbarsoft(\eta_c,\rs),\ksoft(\eta_c,\rs))
     +\\ (1-\alphasoft(\rs)) 
		 P_n^{\text{PaNB}}(\nbarsemi(\eta_c,\rs),\ksemi(\eta_c,\rs)) ,
  \label{eq:1}
\end{multline}
where $P_n^{\text{PaNB}}(\nbar,k)$ are Pascal(negative binomial) multiplicity
distributions with characteristic parameters $\nbar$, the average charged
multiplicity, and $k$, related to the dispersion $D$ by
$k^{-1} = (D^2-\nbar)/\nbar^2$, in the pseudo-rapidity interval
$|\eta|<\eta_c$  at c.m.\ energy $\sqrt{s}$ for the soft and semi-hard
components, and $\alphasoft$ is the fraction of soft events with respect
to the total number of events.
The most appealing result has been the decrease of the average number
of clans of the semi-hard component $\Nbarsemi(\eta_c)$ at fixed
$\eta_c$ as $\rs$ increases (already seen in FPS \cite{combo:prd})
both in the strong-KNO-scaling-violating scenario (where $\ktot^{-1}
\simeq \ln s$) and in the QCD-inspired one (where $\ksemi^{-1} \simeq
1/\sqrt{\ln s}$), together with the corresponding increase of the
average number of particles per clan. 
In addition, the decrease of $\Nbarsemi(\eta_c)$ was found to be
quicker in the strong-KNO-scaling-violating scenario than in the 
QCD-inspired one and more pronounced in larger intervals than in
smaller ones.

Finally the average number of clans increased with the width of the
rapidity interval; the average number of
particles per clan $\ncsemi$ also increased with the width of the rapidity
interval and, contrary to $\Nbarsemi$, increased also with
c.m.\ energy.

This situation raised intriguing questions, on when the asymptotic
limit $\Nbarsemi\to 1$ will be reached and on the consequences of its
eventual earlier occurrence, say at LHC energy.

In \cite{RU:NewPhysics} we concluded that an early occurrence of the
limit $\Nbarsemi\to 1$ in FPS is of particular interest and leads in our
framework to the onset of a new class of events with quite remarkable
properties controlled by $\kiii < 1$, the effective benchmark of the
new class of events.

In particular are expected:

 a) very strong forward-backward (FB) multiplicity
correlations (the FB multiplicity correlations strength is close to
its maximum, $\bFB\sim 1$) with leakage parameter controlling the flow of
particles from one hemisphere to the opposite one \cite{RU:FB} $p\sim
1/2$, again its maximum value;

 b) an enhancement of two-particle correlations determined by
the very large value of $\kiii^{-1}$;

 c) a characteristic elbow structure in $P_n^{\text{total}}$
for large $n$ and a narrow peak for $n$ close to zero.
Both trends are consequences of the log-convex gamma shape of the
$n$-charged particles multiplicity distribution of the new component
which shows a high peak at very low multiplicities and a very slow
decrease for large ones, a general behaviour to be contrasted with
Pythia Monte Carlo calculation predictions, which show at 14 TeV a
second shoulder to be added to the first one already seen in the GeV
energy domain.

\begin{figure}
  \begin{center}
  \mbox{\includegraphics[width=0.7\textwidth]{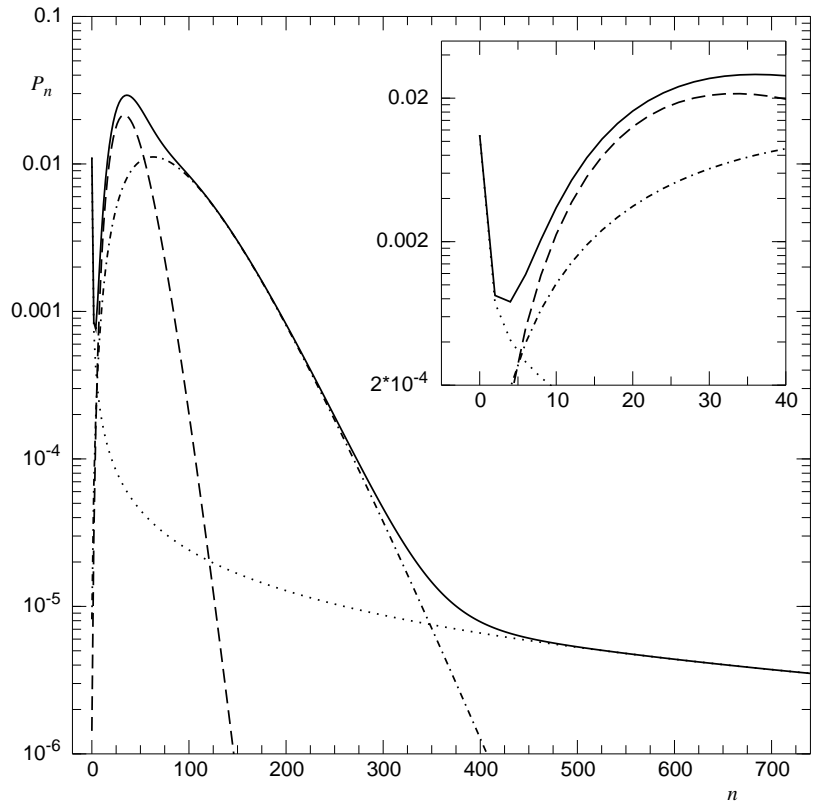}}
  \end{center}
  \caption{Full phase-space multiplicity distribution for the 
    scenario described in the text (solid line); the three 
	components are also shown: soft (dashed line), semi-hard
    (dash-dotted line) and the third (dotted line). 
	The inset shows a magnification of the low-multiplicity peak.}\label{fig:1}
  \end{figure}

\begin{figure}
  \begin{center}
  \mbox{\includegraphics[width=0.7\textwidth]{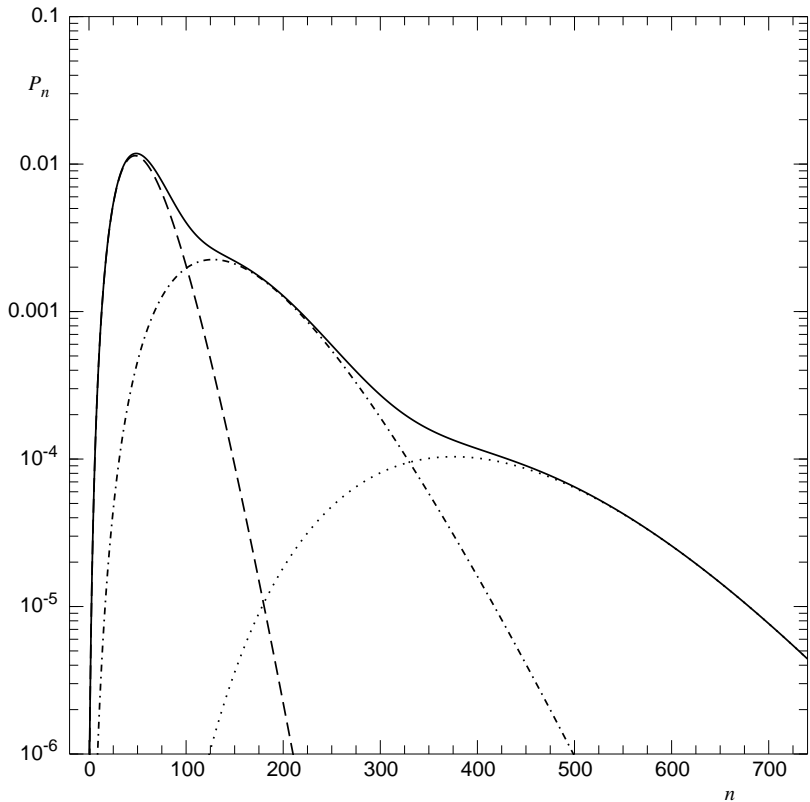}}
  \end{center}
  \caption{Three-component NBD fit to the full phase-space 
  multiplicity distribution predicted by
  Pythia Monte Carlo calculations at 14 TeV; the three component NBD's
  are also shown.}\label{fig:2}
  \end{figure}

\begin{table}
  \caption{\textbf{(a)} Parameters for the extrapolated component MD's 
  at 14 TeV in FPS; 
  \textbf{(b)} same as (a), but in the pseudo-rapidity interval
  $|\eta|<0.9$, with the two extreme scenarios: evenly spread over the whole
  rapidity range (i) and concentrated in the small interval (ii);
  \textbf{(c)} Parameters of NB fits to Pythia predictions in FPS;
  \textbf{(d)} same as (c), but in the pseudo-rapidity interval
  $|\eta|<0.9$.}\label{tab:1}
	\begin{center}
  \vspace*{0.4cm}
  \begin{tabular}[t]{|r|r|ccccc|}
		\hline
		a & FPS\vphantom{\LARGE H} &    \%   &   $\nbar$   &   $k$ &  $\Nbar$ &  $\nc$\\
      \hline
		  & soft    &      41  &   40        &    7     &       13.3    &
		  3.0\\
			& semi-hard  &    57  &   87      &     3.7     &      11.8&
		  7.4\\
			& third      &    2  &  460       &    0.1212    &     1     &
      460\\
			\hline
      \hline
		b& $|\eta|<0.9$\vphantom{\LARGE H} &    \%   &   $\nbar$   &   $k$ &  $\Nbar$ &  $\nc$\\
      \hline
		  & soft    &41 &    4.9     &     3.4       &   3.0      &   1.6\\
			& semi-hard  & 57 &    14  &         2.0   &       4.2  &   3.4\\
			& third (i) & 2  &   40    &       0.056   &     0.368  &  109\\
			& third (ii) & 2  &   460   &       0.1212  &      1     &
      460\\
			\hline
      \hline
		c& FPS\vphantom{\LARGE H} &    \%   &   $\nbar$   &   $k$ &  $\Nbar$ &  $\nc$\\
      \hline
		  & first   &    42 $\pm$ 4  &  52.8 $\pm$ 0.2  & 11 $\pm$ 1  &
      19.3 $\pm$ 1.4 &  2.7 $\pm$ 0.2
		  \\
			& second  &   56 $\pm$ 4  &   123 $\pm$ 3 &   2.7 $\pm$ 0.2 &
      10.5 $\pm$ 0.7 & 11.7 $\pm$ 1.2
		  \\
			& third      &   $\approx 2$  &   468 $\pm$ 7  &   23 $\pm$ 3   &
      70 $\pm$ 7  &   6.6 $\pm$ 0.6\\
			\hline
      \hline
		d& $|\eta|<0.9$\vphantom{\LARGE H} &    \%   &   $\nbar$   &   $k$ &  $\Nbar$ &  $\nc$\\
      \hline
		  & first    &   52 $\pm$ 7  &   5.1 $\pm$ 0.2 & 2.8 $\pm$ 0.3  &
      2.9 $\pm$ 0.1 &  1.75 $\pm$ 0.1
		  \\
			& second  &  47 $\pm$ 7  &  19.1 $\pm$ 1.5 & 1.9 $\pm$ 0.5  &  4.5
      $\pm$ 0.7 &   4.2 $\pm$ 0.7
		  \\
			& third      &  $\approx 1$  &     87.7 $\pm$ 3.7 &  11 $\pm$ 4  &
      24 $\pm$ 5     & 3.6 $\pm$ 0.8\\
			\hline
  \end{tabular}
	\end{center}
  \end{table}

The situation is summarised in Fig.~\ref{fig:1} within the framework of
the weighted superposition mechanism of the three classes of events
and in Fig.~\ref{fig:2} for Pythia Monte Carlo predictions.
It should be noticed that the total MD also in the case of Pythia can
be fitted in terms of the superposition of three NB MD's with good
chi-square per degree of freedom (107/66).
The corresponding characteristic NB parameters of the two cases are
given in Table~\ref{tab:1}a and \ref{tab:1}c respectively.
Striking differences between the behaviours of events of the third
class, seen just by inspection of Fig.s~\ref{fig:1} and \ref{fig:2},
are shown explicitely in the mentioned Tables.

\section{Perspectives: global properties in rapidity intervals}
Tevatron data seem to favour, among our scenarios, the one based on a
strong KNO-scaling violation (i.e., with $\ktot^{-1} \simeq \ln s$). 
In fact it was shown by CDF \cite{CDF:soft-hard}
that, in their data, the component rich in mini-jets
violates KNO scaling in small rapidity intervals, although a cut-off
at low $p_T$ has been used;
in addition, E735 data in FPS \cite{Walker} show again that the
closest scenario is the one mentioned above, although discrepancies
with previous UA5 results \cite{UA5:3}
are noticed in E735 results at lower
c.m.\ energies. 
Therefore, we decided to discuss in rapidity intervals the case of the
strong-KNO-scaling-violating scenario only.

In going from FPS to (pseudo)-rapidity intervals our main concern has
been to be consistent with the scenarios explored in FPS. 
The weight of each component is obviously the same as in FPS.
This leads
to a particle density which shows an energy-independent plateau around
$\eta=0$ in both the soft and the semihard components, a constant
$\ksoft$ value and a linearly increasing $\ktotal$.
It should be remarked that the plateau of the semihard
component is less wide than that of the soft component, but it
is higher. See \cite{combo:eta} for a detailed discussion on these points.

\begin{figure}
  \begin{center}
  \mbox{\includegraphics[width=0.7\textwidth]{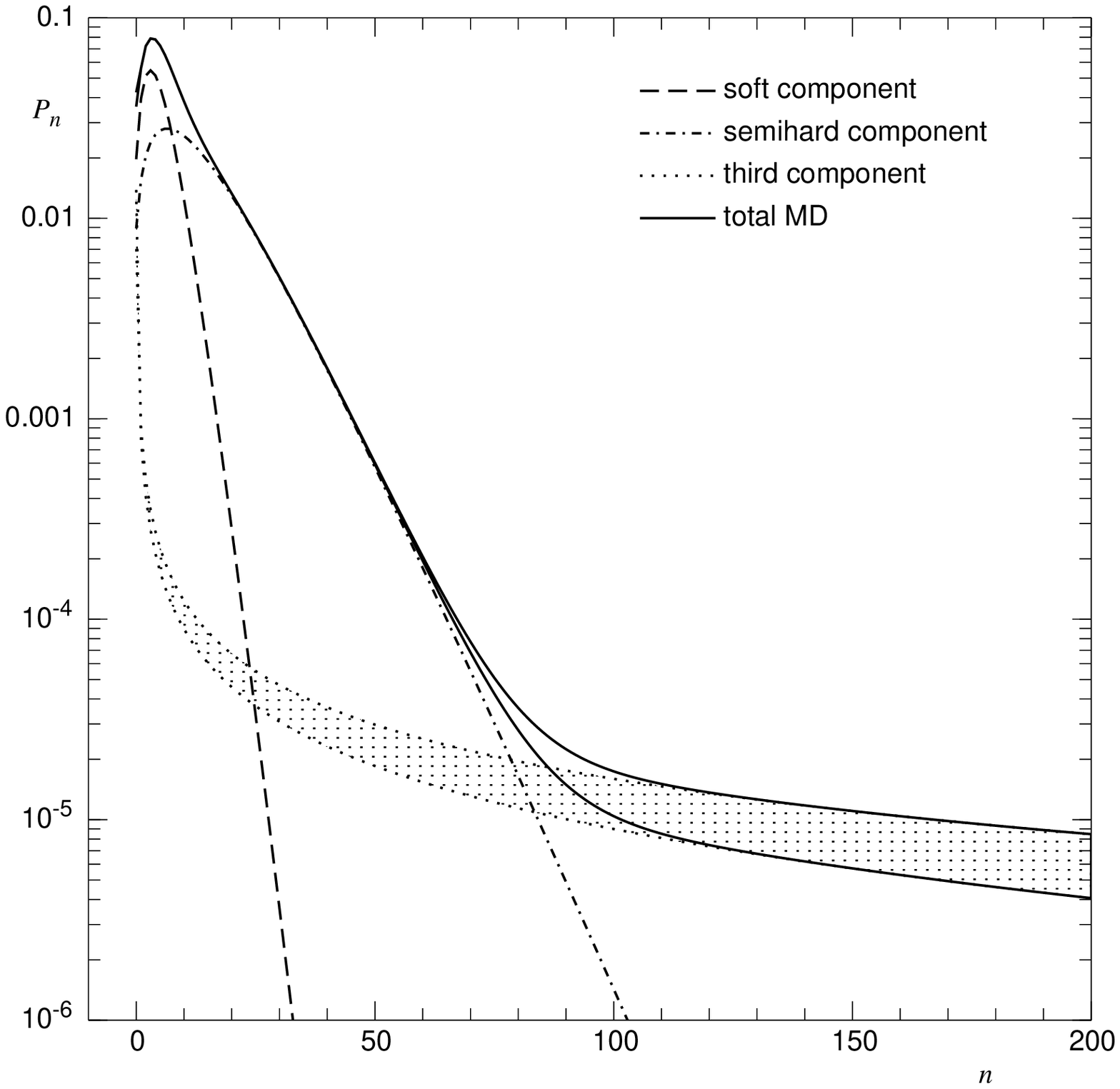}}
  \end{center}
  \caption{Multiplicity distribution in $|\eta|<0.9$ for the 
    scenario described in the text (solid line); the three 
	components are also shown: soft (dashed line), semi-hard
    (dash-dotted line) and the third (dotted line).}\label{fig:3}
  \end{figure}

For the third component, we have  allowed for two extreme behaviours: 
(i) the third
component is distributed uniformly over the whole of phase space and 
(ii) the third component has a very narrow 
and tall plateau
and falls entirely within the interval $|\eta|<0.9$.
These two extremes are represented as a band in the figure.
In the first case, the value of $\kiii$ has again been
determined from the asymptotic behaviour of the average number of
clans in the second (semihard) component, where it is a fixed fraction
of the same quantity in FPS.
Results are shown in Figure~\ref{fig:3} and Table~\ref{tab:1}b.

\begin{figure}
  \begin{center}
  \mbox{\includegraphics[width=0.7\textwidth]{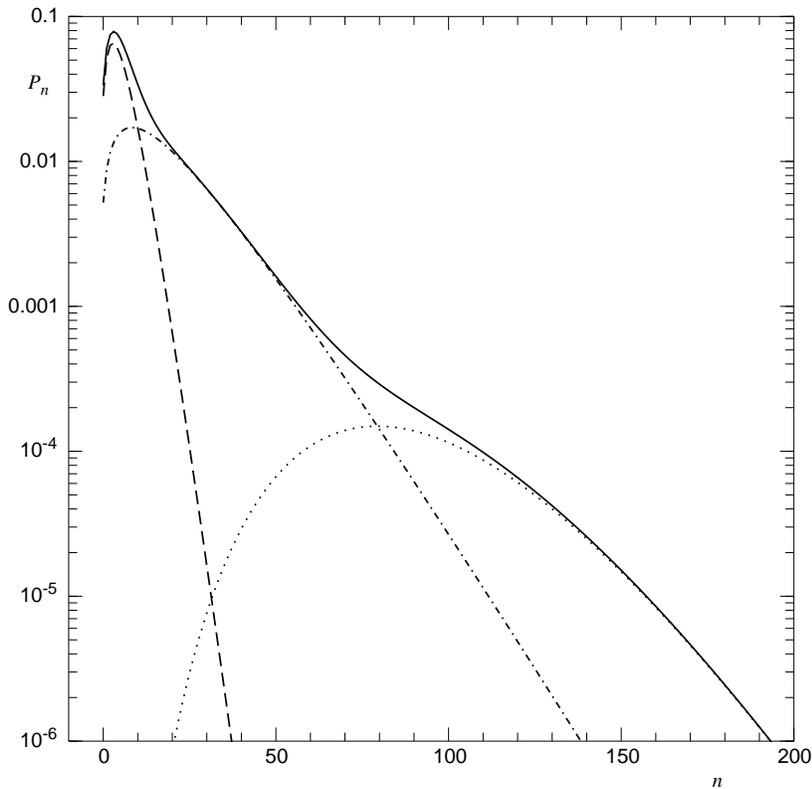}}
  \end{center}
  \caption{Three-component NBD fit to the 
  multiplicity distribution in $|\eta|<0.9$ predicted by
  Pythia Monte Carlo calculations at 14 TeV; the three component NBD's
  are also shown.}\label{fig:4}
  \end{figure}

\begin{table}
  \caption{Forward-backward multiplicity correlation strength
  both in our scenarios and in Pythia Monte Carlo calculations.}\label{tab:2}
  \vspace*{0.4cm}
	\begin{center}
  \begin{tabular}{|r|cc|ccl|}
		\hline
		    &\multicolumn{2}{c}{Pythia}\vline &\multicolumn{2}{c}{Our scenarios}&\\
                     &    FPS   & $|\eta|<0.9$  &    FPS    &
                     $|\eta|<0.9$ & \\
    \hline
    first/soft       &    0.39  &     0.28      &    0.41   &     0.25 & \\
    second/semihard  &    0.53  &     0.48      &    0.51   &     0.45 & \\
    third            &    0.91  &     0.80      &    0.9995 &     0.997 & (i)\\
                     &          &               &           &     0.9995& (ii)\\
    total (weighted) &    0.75  &     0.69      &    0.98   &
		    0.92 & \\
		\hline
  \end{tabular}
	\end{center}
  \end{table}

It should be pointed out that the general trend of $P_n$ vs.\ $n$ is
quite similar to that already seen in FPS: the elbow structure is
clearly visible in both extreme cases; on the other hand, the narrow 
peak at very low multiplicities (also due to the third component) is
hidden by the standard peaks of the soft and semi-hard components
which are shifted to lower multiplicities than in FPS.

This behaviour is clearly seen in Tables~\ref{tab:1}a and \ref{tab:1}b 
by comparing $\nbarsemi$
with $\nbarsoft$ and $\ksemi$ with $\ksoft$ in FPS and in rapidity
intervals.
In addition it turns out that $\Nbarsoft(\FPS) $ is slightly larger
than $ \Nbarsemi(\FPS)$ and
$ \Nbarsoft(|\eta|<0.9)$ is slightly smaller than $\Nbarsemi(|\eta|<0.9)$,
but the average number of particles per clan is both in FPS and in
$|\eta|<0.9$ much larger in the semi-hard than in the soft component,
a fact which confirms the general trend that semi-hard clans are
larger than soft clans.

As far as the third component is concerned, we have the mentioned
extreme situations according to our assumptions:

 i) a situation in which the single clan is uniformly spread
over the whole of phase space (in this case only 37\% of the clan is
contained within the pseudo-rapidity interval $|\eta|<0.9$), $\kiii$ is
even much less than 1 ($\kiii\sim 0.056$) and $\nbariii \sim 40$.

 ii) a situation in which the single clan is fully contained
in $|\eta|<0.9$, i.e., its characteristic parameters are the same as
those seen in FPS.
It is quite clear that particle density in rapidity in this second
case is much higher than in case (i).

In Fig.~\ref{fig:4} are shown the three components and their
superposition predicted by Pythia at 14~TeV in the interval
$|\eta|<0.9$.
We have taken the results from \cite{RU:NewPhysics}, 
where Pythia version 6.210 \cite{Pythia} was
run with default parameters using model 4 with a double Gaussian
matter distribution. This set-up gives a reasonable description of lower
energy MD data.
The same events where then analysed both in FPS and in
the rapidity interval $|\eta| < 0.9$.
The MD's thus found have been fitted with the weighted sum of three
NBD's (it was not possible to use just two NBD's), which have been
taken as the three components. The parameters of the NBD fits can be
found in Table~\ref{tab:1}d ($\chi^2/\text{NDF}=34/42$).

As in FPS, the behaviour of the tail is completely different from that
predicted by our scenarios. It should be noticed that the average
number of particle per clan is quite small in all components although
the average number of clans increases from the first to the third
component suggesting Poissonian MD structure in each component.

Coming to FB multiplicity correlations, it is important to stress that
our previous calculations \cite{RU:FB} of the correlation strength,
$\bFB$, for the two components,
\begin{equation}
  \bFB = \frac{\alpha_1 \frac{b_1 {D^2_{1}}}{1+b_1} +
        (1-\alpha_1) \frac{b_2 {D^2_{2}}}{1+b_2} +
                \frac{1}{2}\alpha_1(1-\alpha_1)(\nbar_{2} - 
\nbar_{1})^2}
              {\alpha_1  \frac{D^2_{1}}{1+b_1} +
                (1-\alpha_1)  \frac{D^2_{2}}{1+b_2} +
             \frac{1}{2}\alpha_1(1-\alpha_1)(\nbar_{2} -
                                \nbar_{1})^2}  ,
\end{equation}
where, for the $i$-th component, $\nbar_i$ is its average
multiplicity, $D_i$ its dispersion, $b_i$ its FB correlation strength
and $\alpha_i$ its weight,
should be extended to accommodate the third one. 
Accordingly, we derived the following
formula for the overall strength for an arbitrary number $M$ of components:
\begin{equation}
	\bFB = \frac{
		\sum_{i=1}^M \alpha_i \frac{b_i D^2_i}{1+b_i} + 
       \frac12 \sum_{i=1}^M \sum_{j>i}^M \alpha_i\alpha_j(\nbar_i - \nbar_j)^2
  }{
		\sum_{i=1}^M \alpha_i \frac{D^2_i}{1+b_i} + 
       \frac12 \sum_{i=1}^M \sum_{j>i}^M \alpha_i\alpha_j(\nbar_i - \nbar_j)^2
  } .
	\label{eq:2}
\end{equation}
Applied to the three components case, Eq~(\ref{eq:2}) leads to the
results shown in Table~\ref{tab:2}, where the leakage parameter in rapidity
intervals has been taken to be the same as in FPS \cite{RU:FB}.
In particular it is clear that the FB correlation strength for the
soft component is quite larger in FPS than in $|\eta|<0.9$
(${\bFB}_{,\text{soft}}(\FPS) = 0.41$, 
${\bFB}_{,\text{soft}}(|\eta|<0.9) = 0.25$)
and still remains larger in FPS than in $|\eta|<0.9$ for the semi-hard
component 
(${\bFB}_{,\text{sh}}(\FPS) = 0.51$, 
${\bFB}_{,\text{sh}}(|\eta|<0.9) = 0.45$),
although it is always much larger both in FPS and in $|\eta|<0.9$ for
the semi-hard component than for the soft one.
The third component tends to saturate in all cases its maximum value,
which is 1.
The total FB multiplicity correlation strength resulting from the
composition of the contributions of all classes of events is larger in
FPS ($\bFB = 0.98$) than in $|\eta|<0.9$ ($\bFB = 0.92$) but closer to
its asymptotic value.

FB correlations in Pythia do not show remarkable differences with
respect to our scenarios except in the third component and in the
total MD where they are considerably smaller; in particular the
difference with the third component is striking (0.8 Pythia, $\approx
1$ our scenarios).
Stronger FB correlations at hadron level suggest an extraordinary
stronger colour exchange process at parton level in the last case.

\begin{table}
  \caption{Energy density and corresponding parameters for our scenarios and for
  Pythia Monte Carlo. The volume $V=\pi R^2\tau$ has been computed with
  proton radius $R\approx 1.1$~fm and formation time 
  $\tau\approx 1$~fm.}\label{tab:Bj}
	\begin{center}
  \begin{tabular}[t]{|r|r|cccccc|}
		\hline
		 a& our scenarios& soft & semi-hard & third (i) &  (ii) & 
		   total (i) & (ii)\\
    \hline
		&$dn/dy$ &  2.5 & 7 & 20 & 230 & 10.8 & 19.2 \\
		&$\avg{E_T}$ (MeV)&  350 & 500 & 500 & 500 & 500 & 500 \\
		&$\varepsilon$ (GeV/fm${}^3$)& 0.41 & 1.6 & 4.7 & 54 & 2.5 & 4.5\\
		\hline
		\hline
		 b& Pythia & first & second & \multicolumn{2}{c}{third} &  
		  \multicolumn{2}{c}{total} \vline\\
    \hline
		&$dn/dy$ & 2.5 & 9.5 & \multicolumn{2}{c}{44} & 
		\multicolumn{2}{c}{12.5}\vline\\
		&$\avg{E_T}$ (MeV)& 350 & 500 & \multicolumn{2}{c}{500} &
		\multicolumn{2}{c}{500} \vline\\
		&$\varepsilon$ (GeV/fm${}^3$)&  0.41 & 2.4 &
		 \multicolumn{2}{c}{10} & \multicolumn{2}{c}{3.0} \vline\\
		\hline
  \end{tabular}
	\end{center}
  \end{table}

In addition, Bjorken formula \cite{Bj:energy} for the energy density,
\begin{equation}
	\varepsilon = \frac32 \frac{\avg{E_T}}{V} 
	        \left.\frac{dn}{dy}\right\vert_{y=0},
\end{equation}
where $\avg{E_T}$ is the average transverse energy per particle, $V$
the collision volume and $dn/dy$ the particle density at mid-rapidity,
has been applied in order to compare its predictions on pp collisions
with those on nucleus-nucleus collisions.
Parameters of the formula and results are shown in Table~\ref{tab:Bj}a:
$dn/dy$ values correspond to predictions for our scenarios already
seen in Table~\ref{tab:1}; lacking general expectations for the
average transverse energy $\avg{E_T}$, we used for the soft component
the value measured at ISR and, in a conservative way, the value
measured by CDF for the other components (to be intended as a lower
bound, which leads to lower bounds for the energy density as well).

It should be noticed that:

a) the energy density for the semi-hard component in our scenario at 
14~TeV is $\approx 1.6$, i.e., of the same order of magnitude of that
found at AGS at 5.6~GeV in O+Cu collisions ($\varepsilon\approx 1.7$).

b) the energy density for the third component in the spread out
scenario is 4.7, i.e., comparable with the value of $\varepsilon$
recently measured at RHIC in Au+Au collisions ($\varepsilon\approx 4.6$).

c) the energy density for the third component in the other extreme
scenario (high concentration) is $\approx 54$ even larger, being $dn/dy$ much 
larger, than the LHC
expectations for central Pb+Pb collisions ($\varepsilon\gtrsim 15$).

On the other side, 
Pythia prediction for its third component (see Table~\ref{tab:Bj}b) is
contained within our extreme predictions, and it is of the same order of
magnitude in the other components.
Notice that the full minimum bias sample is also intermediate between
our scenarios results.

Of course our calculation of $\varepsilon$ is only indicative and
should be taken with caution.
Although the use of Bjorken formula for pp collisions as well as the
choice of parameters is rather doubtful, we consider our results quite
stimulating because they suggest the possibility that the same
characteristic behaviour of many observables seen at RHIC energies in
AA collisions could be reproduced at LHC in pp collisions.

\bigskip
The implications of the existence of a third class of events
in minimum bias pp collisions characterised by the presence of one or  
very few clans, i.e., by the
corresponding NB parameter  $\kiii < 1$, has been examined in pseudo-rapidity 
intervals accessible at LHC ($|\eta| < 0.9$)  paying attention to
total charged particle  multiplicity distributions, forward-backward
multiplicity correlations and energy density.
Assuming that the  guesswork at the basis of our calculations will be 
confirmed, more than try to draw conclusions one should ask two intriguing 
questions for future experimental and theoretical work:

1. what new physical phenomenon is hidden behind the 
squeezing of one or very few clans 
in the rapidity interval $|\eta|< 0.9$ with much stronger 
forward-backward multiplicity correlations than the soft and semi-hard 
components, and  with energy density  comparable with that of 
nucleus-nucleus at RHIC?

2. what would be the QCD counterpart of this new phenomenon? would it
   be a signal of parton saturation?


\bibliographystyle{prstyR} 
\bibliography{abbrevs,bibliography}

\end{document}